\documentclass[pra,twocolumn,showpacs,preprintnumbers,amsmath,amssymb,tightenlines,epsfig,floatfix]{revtex4}

\usepackage{graphicx}% Include figure files
\usepackage{dcolumn}% Align table columns on decimal point
\usepackage{bm}% bold math
\usepackage{amsfonts}

\begin{document}
\preprint{Version: submitted \today}
\title{Outcoupling from a Bose-Einstein condensate with squeezed light to produce entangled atom laser beams. }
\author{S.A. Haine and J.J. Hope}
\affiliation{Australian Centre for Quantum-Atom Optics, The Australian National University, Canberra, 0200, Australia.}

\email{simon.haine@anu.edu.au}

\begin{abstract}
We examine the properties of an atom laser produced by outcoupling from a Bose-Einstein condensate with squeezed light. We model the multimode dynamics of the output field and show that a significant amount of squeezing can be transfered from an optical mode to a propagating atom laser beam.  We use this to demonstrate that two-mode squeezing can be used to produce twin atom laser beams with continuous variable entanglement in amplitude and phase. 
\end{abstract}


\maketitle
\section{Introduction}

The experimental demonstration of Bose-Einstein condensates (BEC) \cite{dalfovo} has lead to the development of atom lasers  by outcoupling atoms from trapped BECs by either a radio frequency transition or a Raman transition to change the internal state of the atom to one that is either untrapped or anti-trapped \cite{andrews, mewes, martin, bloch, hagely, anderson, lecoq, Cennini, Robins}. Atom lasers are coherent matter waves with spectral fluxes many orders of magnitude higher than thermal sources of atoms.  The coherence of these sources will enable an increase in the sensitivity of interferometric measurements \cite{atomint}.  Although current experiments usually operate in parameter regimes limited by technical noise, the fundamental limit on these measurements will be caused by the shot noise of the atomic field, which will be intrinsic to all interferometers without a non-classical atomic source.  Sensitivity is increased in optical interferometry by `squeezing' the quantum state of the optical field, where the quantum fluctuations in one quadrature are reduced compared to a coherent state, while the fluctuations in the conjugate quadrature are increased.  In the context of atom optics, it is interesting to ask whether highly squeezed atom optical sources can be produced.  There is also great interest in the production of entangled atomic beams for quantum information processing and tests of quantum mechanics with massive particles \cite{entangledinterest}.  This paper will describe methods of coupling the quantum statistics from optical fields to produce non-classical atomic sources with high efficiency.

Generation of squeezed atomic beams has been proposed by either utilising the nonlinear atomic interactions to create correlated pairs of atoms via either molecular down conversion or spin exchange collisions \cite{Duan1, Pu, Kheruntsyan}, or by transferring the quantum state of a squeezed optical field to the atomic beam \cite{Moore, Jing, Fleischhauer}. In the first case, it was shown that collisions between two condensate atoms in the $|M_F = 0\rangle$ state can produce one atom in the $|M_F = +1\rangle$ and one atom in the $|M_F = -1\rangle$, with sufficient kinetic energy to escape the trap \cite{Duan1,Pu}. It was shown that this scheme produced pairs of atoms entangled in atomic spin. In the second scheme a BEC of molecules composed of bosonic atoms is disassociated to produce twin atomic beams, analogous to optical down conversion \cite{Kheruntsyan}. It was shown that the beams were entangled in the sense that phase and amplitude measurements on one beam could infer phase and amplitude measurements of the other beam better than the Heisenberg limit.  Although each atomic pair is perfectly correlated in direction in each of these schemes, there is very little control of direction of each pair, so the spectral flux would be limited. 

The generation of nonclassical light is well established experimentally \cite{Bachor}. This suggests that a nonclassical atom laser output could be generated by transferring a the quantum state of an optical mode to an atomic beam. Moore {\it et al.} showed that a quantized probe field could be partially transferred to momentum `side modes' of a condensate consisting of three-level atoms in the presence of a strong pump field \cite{Moore}. Jing {\it et al.} performed a single mode analysis of the atom laser outcoupling process for a two-level atom interacting with a quantized light field, and showed that the squeezing in light field would oscillate between the light field and the atomic field at the Rabi frequency \cite{Jing}. As this was a single mode analysis, the interaction with the atoms as they left the outcoupling region was not taken into account. Fleischhauer {\it et al.} \cite{Fleischhauer} showed that Raman adiabatic transfer can be used to transfer the quantum statistics of a propagating light field to a continuously propagating beam of atoms by creating a polariton with a spatially dependent mixing angle, such that the output contained the state of the probe beam. 

In this paper, we model the dynamics of an atom laser produced by outcoupling three-level atoms from a BEC via a Raman transition, and investigate the transfer of quantum statistics from one of the optical modes to the atomic field.  Ideal transfer will occur when the time taken for each atom to leave the outcoupling region is a quarter of a Rabi period.  The finite momentum spread of a trapped condensate means that there will be a broadening of the time taken to leave the outcoupling region, and hence ideal transfer will not be possible. To determine the effectiveness of the quantum state transfer, we require a multimode model that takes into account back coupling and the finite momentum spread of the condensate. 

In Section II we describe an atom laser beam made by outcoupling from a BEC using a non-trivial optical mode, using the simplest possible model that contains the spatial effects in the output mode.  We derive the Heisenberg equations of motion for this system under suitable approximations.  Section III introduces the method used to solve these equations and investigates some properties of the outcoupled atoms, showing that complicated spatial behaviour occurs in the output even when the optical and BEC fields are described by a single mode.  In section IV we investigate continuous outcoupling with two mode squeezing, and show that it can be used to generate continuous variable entanglement in twin atom laser beams propagating in different directions. 

\section{Outcoupling using a nonclassical optical field}

When an atomic and an optical field are coupled, and they can both be described by a single mode, then complete state transfer must occur between them in a Rabi-like cycle.  When producing an atom laser beam in this manner, however, the single mode approximation cannot be made for the output field, even though it may be applicable to the optical and BEC fields.  In this section we develop such a model, and derive the Heisenberg equations of motion for the output field and the optical field operators.

\begin{figure}
\includegraphics[width=\columnwidth]{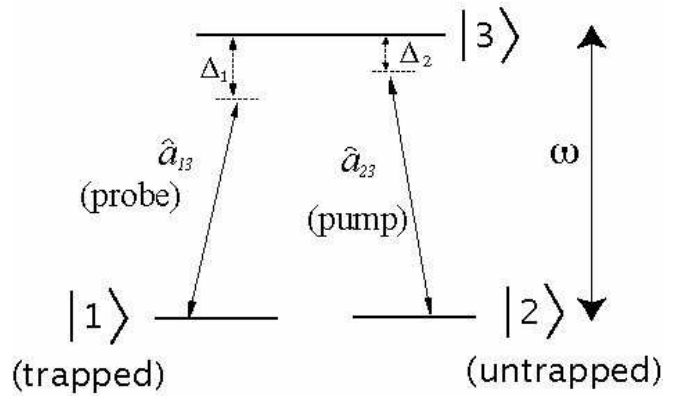}
\caption{\label{fig:levels} Internal energy levels of a three level atom. A condensate of state $|1\rangle$ atoms confined in a trapping potential are coupled to free space via a Raman transition affected by a probe beam (annihilation operator $\hat{a}_{13}$) which is detuned from the excited state ($|3\rangle$) by an amount $\Delta_1$, and a pump field (annihilation operator $\hat{a}_{23}$) which is detuned from the excited state by an amount $\Delta_2$.}
\end{figure}

We model an atom laser in one dimension as a BEC of three-level atoms coupled to free space via a Raman transition, as shown in figure \ref{fig:levels}.  State $|1\rangle$ represents the internal state of the trapped condensate, $|3\rangle$ the excited state, and $|2\rangle$ the untrapped atomic mode. $\hat{a}_{13}$ is the annihilation operator for the probe optical mode (transition $|1\rangle \rightarrow |3\rangle$), and  $\hat{a}_{23}$ is the annihilation operator for the pump optical mode (transition $|2\rangle \rightarrow |3\rangle$). The pump field is assumed to be a large coherent state, much stronger than the probe field, so it is approximated well by a classical field $g_{23}\hat{a}_{23} = \Omega^{*}_{23}e^{-i(\omega-\Delta_2)t}$. 
The Hamiltonian for the system (in the rotating wave approximation) is: 
\begin{eqnarray}
\hat{\mathcal H} &=& \hat{\mathcal H}_{atom} + \hat{\mathcal H}_{light} + \hat{\mathcal H}_{atom-light} \\ \nonumber 
&=&  \int \hat{\psi}^{\dag}_1(k) H_0 \hat{\psi}_1(k) dk +\frac{\hbar^2}{2m} \int k^2 \hat{\psi}^{\dag}_2(k)\hat{\psi}_2(k) dk \\ \nonumber
&+&  \int \hat{\psi}^{\dag}_3(k)(\frac{\hbar^2k^2}{2m} +\hbar \omega)\hat{\psi}_3(k) dk
+ \hbar(\omega -\Delta_1)\hat{a}_{13}^{\dag}\hat{a}_{13} \\ \nonumber
&+& \hbar g_{13} \int \hat{\psi}_1(k)\hat{\psi}_3^{\dag}(k+k_{13})\hat{a}_{13} + \hat{\psi}^{\dag}_1(k)\hat{\psi}_3(k+k_{13})\hat{a}_{13}^{\dag} dk \\ \nonumber
&+& \hbar \int \Omega_{23} \hat{\psi}^{\dag}_2(k)\hat{\psi}_3(k+k_{23})e^{i(\omega-\Delta_2)t} \\ \nonumber
&+&  \Omega^{*}_{23} \hat{\psi}_2(k)\hat{\psi}^{\dag}_3(k+k_{23})e^{-i(\omega-\Delta_2)t} dk \\ \nonumber
\end{eqnarray}
 where $\hat{\psi}_1(k)$ is the k-space annihilation operator the condensate mode (internal state $|1\rangle$), $\hat{\psi}_3(k)$ is the annihilation operator for atoms in the excited atomic state ($|3\rangle$), and $\hat{\psi}_2(k)$ is the annihilation operator for the untrapped free propagating mode ($|2\rangle$). The annihilation operators obey the usual bosonic commutation relations:
 \begin{align}
 [\hat{\psi}_i(k),\quad \hat{\psi}_j(k')] = [\hat{\psi}^{\dag}_i(k),\quad \hat{\psi}^{\dag}_j(k')] =0, \\ \nonumber
  [\hat{\psi}_i(k),\quad \hat{\psi}^{\dag}_j(k')]=\delta_{ij}\delta(k-k') 
 \end{align}
$H_0$ is the single particle Hamiltonian for the trapped atoms, $m$ is the mass of the atoms,  $\Omega_{23}$ is the Rabi frequency for the pump transition, $g_{13}$ is the coupling strength between the atom and the probe field, $\hbar \omega$ is the internal energy of the excited state $|3\rangle$ atoms, and $\hbar k_{13}$ and $\hbar k_{23}$ are the momentum kicks due to the pump and probe light fields respectively. For simplicity we have assumed a laser geometry where the pump and probe fields are counter propagating, to give the maximum possible momentum kick to the untrapped atoms. 
The equations of motion for the Heisenberg operators are:
\begin{eqnarray}
i\dot{\hat{\psi}}_1(k) &=& \frac{H_0}{\hbar}\hat{\psi}_1(k) + g \tilde{\psi}_3(k+k_{13})\hat{a}^{\dag} \\
i\dot{\hat{\psi}}_2(k) &=& \frac{\hbar k^2}{2m}\hat{\psi}_2(k) + \Omega_{23} \tilde{\psi}_3(k+k_{23}) \\
i\dot{\tilde{\psi}}_3(k) &=& ( \frac{\hbar k^2}{2m}+\Delta_2)\tilde{\psi}_3(k) + g_{13} \hat{\psi}_1(k-k_{13})\hat{a}\\ \nonumber
 &+& \Omega^{*}_{23}\hat{\psi}_2(k-k_{23})  \\
i\dot{\hat{a}} &=& \delta\hat{a} + g_{13}\int \hat{\psi}^{\dag}_1(k-k_{13})\tilde{\psi}_3(k)dk
\end{eqnarray}
Where $\tilde{\psi}_3 = \hat{\psi}_3 e^{i(\omega-\Delta_2)t}$ and $\hat{a} = \hat{a}_{13}e^{i(\omega-\Delta_2)t}$, and $\delta = (\Delta_2 - \Delta_1)$ is the two-photon detuning.

The population of state $|3\rangle$ will be much less than the other levels when the detunings ($\Delta_1$, $\Delta_2$) are much larger than the other terms in the system (including the kinetic energy of the excited state atoms). Furthermore, most of the dynamics will occur on time-scales less than $\frac{1}{\Delta_2}$, so in this regime we can set $\tilde{\psi}_3(k, t) \approx \frac{-1}{\Delta_2}(g_{13} \hat{\psi}_1(k-k_{13}, t)\hat{a} + \Omega_{23}^{*}\hat{\psi}_2(k-k_{13}, t))$. If the condensate has a large number of atoms and is approximately in a coherent state, we can write $\hat{\psi}_1(k, t) \approx \sqrt{N}\phi_0(k)e^{-i\omega_{t}t}$, where $\phi_0(k)$ is the condensate wavefunction (which we will assume is in the ground state of the harmonic oscillator, with $\omega_{t}$ the trapping frequency) and $N$ is the condensate number. We have ignored the atom-atom interactions in our model, which is valid only if the condensate is dilute.  Strong atom-atom interactions would have the effect of introducing complicated evolution to the quantum state of the condensate mode.  Inclusion of these effects is not possible with our method, and a more complicated technique such as a phase space method would be required \cite{phase space method}. The approximation of ignoring the back action on the condensate is only valid if we are in the regime where the outcoupling is weak, ie. the number of photons in the probe field is is much less than the number of atoms in the condensate. In an experiment, measuring the quantum noise on the atom laser beam would require small classical noise on the beam, and in practice this is easier to achieve with weak outcoupling. With these approximations our equations of motion for the free propagating atoms and the probe field become
\begin{eqnarray}
i\dot{\hat{\psi}}(k) &=& \omega_0(k)\hat{\psi}(k) - \Omega_0(k)\hat{a}  \label{psidot} \\
i\dot{\hat{a}} &=& \omega_a \hat{a}  - \int \Omega_0^{*}(k)\hat{\psi}(k)dk \label{adot}
\end{eqnarray}

with $\hat{\psi}(k) = \hat{\psi}_2(k)e^{i\omega_{t}t}$, $\omega_0(k) = (\frac{\hbar k^2}{2m} -\frac{|\Omega_{23}|^2}{\Delta_2} -\omega_{t})$, $\omega_a = (\delta -\frac{g_{13}^2N}{\Delta_2}$), and $\Omega_0(k) = g_{13}\sqrt{N} \frac{\Omega_{23}}{\Delta_2} \phi_0(k+k_{23}-k_{13})$.

In the next section we will discuss the solution to these equations and the properties of the outcoupled atoms. 

\section{Properties of the outcoupled atoms}

The solution to equations (\ref{psidot}) and (\ref{adot}) is
\begin{eqnarray}
\hat{\psi}(k,t) &=& \int f(k,k',t)\hat{\psi}_s(k)dk' +g(k,t)\hat{a}_{s} \label{solutionpsi} \\
\hat{a}(t) &=& p(t)\hat{a}_s + \int q(k',t)\hat{\psi}_s(k')dk' \label{solutiona}
\end{eqnarray}
Where $\hat{a}_s = \hat{a}(t=0)$ and $\hat{\psi}_s(k)  = \hat{\psi}(k, t=0)$ are the Schr\"{o}dinger picture operators, and $f(k, k', t)$, $g(k, t)$, $p(t)$, $q(k',t)$ are complex functions satisfying:
\begin{eqnarray}
i\dot{f}(k,k') &=& \omega_0(k)f(k,k') -\Omega_0(k)q(k')  \label{semiclassical} \\ \nonumber
i\dot{g}(k) &=& \omega_0(k)g(k) -\Omega_0(k)p \\ \nonumber
i\dot{p} &=& \omega_a p -\int\Omega^*_0(k)g(k) dk \\ \nonumber
i\dot{q}(k') &=& \omega_a q(k') -\int\Omega^*_0(k)f(k,k') dk \\ \nonumber
\end{eqnarray}
with initial conditions $f(k, k',t=0)= \delta(k-k')$, $p(t=0) = 1$, and $g(k, t=0) = q(k',t=0) =0$.  This ansatz has reduced the field operator equations to a set of coupled partial differential equations.  This will only be possible for Heisenberg equations of motion that do not have terms with products of operators, but it allows the possibility of an analytic or numerical solutions to the full quantum problem.

We solved equations (\ref{semiclassical}) numerically using a fourth order Runge Kutta algorithm using the XMDS numerical pacakge \cite{xmds}. We chose parameters realistic to atoms optics experiments with Rb$^{87}$ atoms. Unless stated otherwise, we have set $m = 1.4\times 10^{-25}$ kg, $\omega_{t} = 0.25$ rad $s^{-1}$, $|{\bf k}_{23} -{\bf k}_{13}| = 1.6\times10^7$ m$^{-1}$, which corresponds to twice the wave number of the $^2$S$_{\frac{1}{2}}$,F$=2$ $\rightarrow$ $^2$P$_{\frac{3}{2}}$,F$=3$ transition in Rb$^{87}$. $\phi_0(k)$ was chosen to be the (normalized) ground state momentum space wave function of a condensate (ignoring interactions) in a  harmonic trap, and we set $\Omega_0(k) = \Omega\phi_0(k - k_{23} -k_{13})$ with $\Omega = 90$ rad s$^{-1}$. The results are reasonably insensitive to the absolute magnitude of $\omega_a$ and $\omega_0$, but they are quite sensitive to the relative values. To maintain resonance between the two fields, we set $\frac{|\Omega_{23}|^2}{\Delta_2} = \frac{\hbar(k_{23} -k_{13})^2}{2m} -\omega_a -\omega_{t}$. These relationships can be obtained with physically realistic parameters and are consistent with the approximations made in this model. We set $\omega_a=20$ rad s$^{-1}$. 

\begin{figure}
\includegraphics[width=\columnwidth, bb=0 0 600 600]{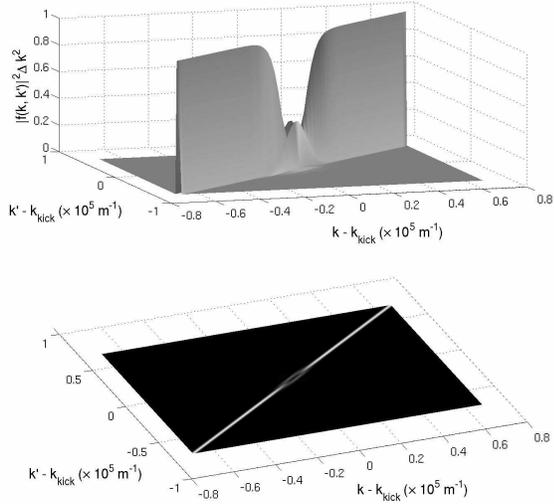}
\caption{\label{fig:f1} $|f(k, k', t=0.11 s)|^2$ for the values of parameters indicated in the text. $k_{kick}$ is the kick acquired due to the Raman transition, ie $k_{kick} = k_{23} - k_{13}$. The function was discretized for numerical calculation by replacing $\delta(k-k')$ with $\frac{\delta_{k, k'}}{\Delta k}$, where $\Delta k$ is the grid spacing. The dip in the function near the $k=k_{kick}$ resonance shows how the quantum state of the atoms has been affected by the interaction with the optical fields.}
\end{figure}

\begin{figure}
\includegraphics[width=\columnwidth, bb=0 0 600 600]{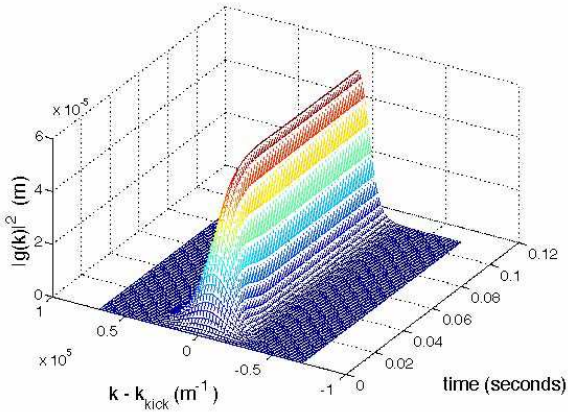}
\caption{\label{fig:g1} $|g(k, t)|^2$ as found numerically for the values of parameters indicated in the text. This shows that atoms are created around $k=k_{kick}$ with a quantum state related to the initial state of the probe field.}
\end{figure}

\begin{figure}
\includegraphics[width=\columnwidth, bb=0 0 600 600]{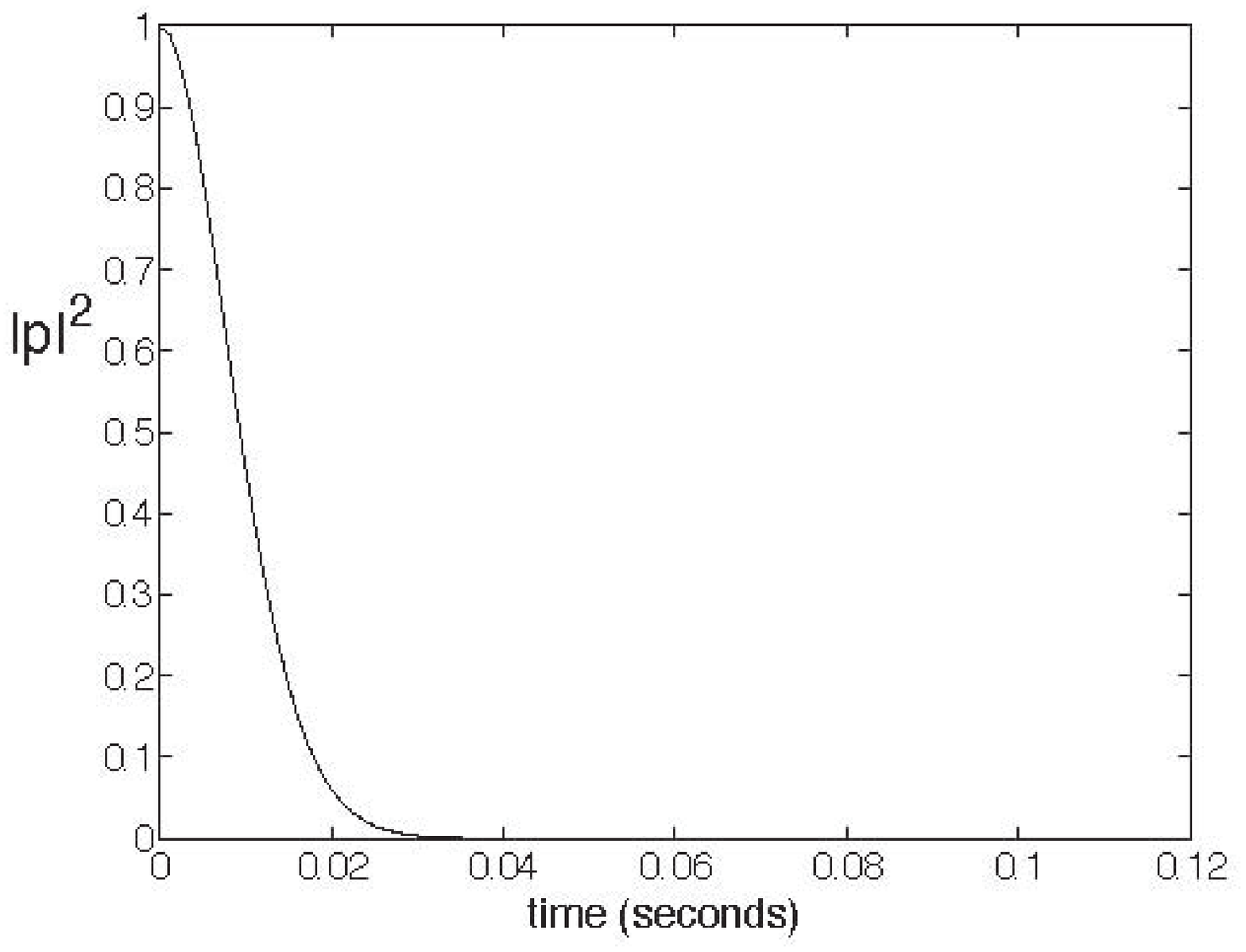}
\caption{\label{fig:p1} $|p(t)|^2$ as found numerically for the values of parameters indicated in the text.}
\end{figure}

\begin{figure}
\includegraphics[width=\columnwidth, bb=0 0 600 600]{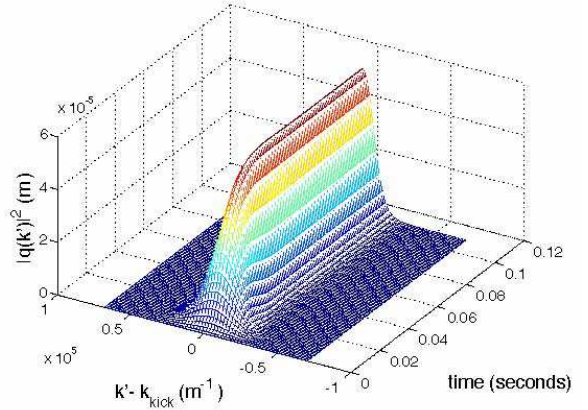}
\caption{\label{fig:q1} $|q(k', t)|^2$ as found numerically for the values of parameters indicated in the text.}
\end{figure}

Figures (\ref{fig:f1}), (\ref{fig:g1}), (\ref{fig:p1}) and (\ref{fig:q1}) show the solutions to equations (\ref{semiclassical}) for the values indicated above. 


\begin{figure}
\includegraphics[width=\columnwidth, bb=0 0 600 600]{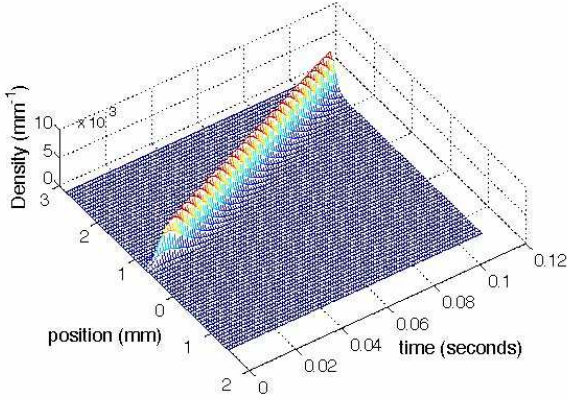}
\caption{\label{fig:xdens} $\langle\hat{\Psi}^{\dag}(x)\hat{\Psi}(x)\rangle$ for $\langle \hat{a}_s^{\dag}\hat{a}_{s}\rangle = 1000$.}
\end{figure}

The solution of equations (\ref{semiclassical}) gives us the solution of Eqs.(\ref{solutionpsi}) and (\ref{solutiona}) for all possible initial quantum states of the optical field and the free propagating atomic field.  We will assume that the initial state of the field is  $|\psi\rangle \equiv |light\rangle \otimes\{|0\rangle\}_k$ \label{initstate}, where $|light\rangle$ represents an arbitrary state for the optical mode, and $\{|0\rangle\}_k$ represents a vacuum mode at all points in $k$ space for the atomic field. The expectation value of the density of outcoupled atoms $\langle \hat{\Psi}^{\dag}(x) \hat{\Psi}(x)\rangle$ with $\hat{\Psi}(x) = \frac{1}{\sqrt{2\pi}} \int \hat{\psi}(k) e^{ikx}dk$ is 
\begin{equation}
\langle \hat{\Psi}^{\dag}(x) \hat{\Psi}(x)\rangle = |G(x)|^2 \langle \hat{a}^{\dag}_s\hat{a}_s\rangle, \quad G(x) = \frac{1}{\sqrt{2\pi}} \int g(k) e^{ikx}dk
\end{equation}
It is interesting to note that when the initial state of the untrapped atomic field is the vacuum, then the spatial structure of the density of the untrapped atoms at later times depends only on the the functional form of $G(x)$, which depends on the efficiency of the outcoupling process.  
Figure (\ref{fig:xdens}) show the density of outcoupled atoms when the expectation value of the initial number of photons is $\langle \hat{a}_s^{\dag}\hat{a}_{s}\rangle = 1000$. 

\begin{figure}
\includegraphics[width=\columnwidth, bb=0 0 600 600]{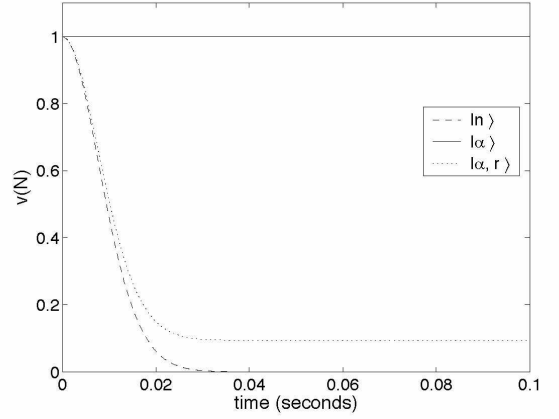}
\caption{\label{fig:Nvar1} The relative variance $v(\hat{N})=\frac{V(\hat{N})}{\langle \hat{N}\rangle}$ for the outcoupled atoms versus time for different states of the optical field. The solid line represents the initial optical mode in a coherent state $|\alpha\rangle$ with $|\alpha|^2 = 1000$, the dotted line represents a squeezed state $|\alpha, r\rangle$ with $|\alpha|^2 = 1000$, $r=1.38$, and the dashed line represents a Fock state $|n\rangle$ with $n=1000$.}
\end{figure}

The number operator is $\hat{N} = \int \hat{\Psi}^{\dag}(x) \hat{\Psi}(x) dx$. Using our solution (Eq. \ref{solutionpsi}) and our initial state, the expectation value is $\langle\hat{N}\rangle =\langle \hat{a}^{\dag}_s \hat{a}_s \rangle \int |G(x)|^2 dx$. The variance of the number operator is
\begin{eqnarray}
V(\hat{N}) &=& \langle \hat{N}^2 \rangle - \langle \hat{N} \rangle^2 \nonumber \\
&=& \int\int \langle \hat{\Psi}^{\dag}(x') \hat{\Psi}(x') \hat{\Psi}^{\dag}(x) \hat{\Psi}(x)\rangle dx dx' \nonumber \\ 
&-&\Big( \int\langle \hat{\Psi}^{\dag}(x) \hat{\Psi}(x) \rangle dx \Big)^2 \nonumber \\
&=& \int \int \langle \hat{\Psi}^{\dag}(x')\hat{\Psi}^{\dag}(x)\hat{\Psi}(x') \hat{\Psi}\rangle(x)dx dx' \nonumber \\
&+&  \int\langle \hat{\Psi}^{\dag}(x) \hat{\Psi}(x) \rangle dx - \Big( \int\langle \hat{\Psi}^{\dag}(x) \hat{\Psi}(x) \rangle dx \Big)^2. \nonumber \\
\end{eqnarray}
Using our solution (Eq. \ref{solutionpsi}) and our initial state, this becomes
\begin{eqnarray}
V(\hat{N}) &=& N_G^2\Big(\langle \hat{a}^{\dag}_s \hat{a}^{\dag}_s \hat{a}_s \hat{a}_s \rangle -\langle  \hat{a}^{\dag}_s \hat{a}_s \rangle^2 \Big) + N_G\langle  \hat{a}^{\dag}_s \hat{a}_s \rangle \nonumber \\
 &=& N_G^2  V(\hat{a}^{\dag}_s \hat{a}_s) + N_G(1-N_G)\langle  \hat{a}^{\dag}_s \hat{a}_s \rangle,
 \end{eqnarray}
 with $N_G = \int |G(x)|^2 dx$. We note that as $N_G \rightarrow 1$, the variance in the number of outcoupled atoms approaches the variance of the initial optical mode, as the quantum statistics of the outcoupled atoms depends only on the initial quantum state of the optical field and the efficiency of the outcoupling process. Figure (\ref{fig:Nvar1}) shows the variance of the outcoupled atoms versus time for different states of the optical mode. 
A more interesting observable to look at is the flux of the outcoupled atoms, as the spatial structure of the outcoupled beam becomes apparent. The flux operator is:
\begin{equation}
\hat{J}(x) = \frac{i\hbar}{2m}\Bigl( \nabla \hat{\Psi}^{\dag}(x) \hat{\Psi}(x) - \hat{\Psi}^{\dag}(x) \nabla \hat{\Psi}(x) \Bigr) 
\end{equation}

which, using our solution for $\hat{\psi}(k)$ becomes

\begin{eqnarray}
\hat{J}(x) &=& \int\int J_f(x,k',k'') \hat{\psi}_s^{\dag}(k')\hat{\psi}_s(k'') dk' dk''  \nonumber \\
 &+& J_g(x) \hat{a}_s^{\dag}\hat{a}_s  \nonumber \\
 &+& \int J_{fg}(x, k') \hat{\psi}_s^{\dag}(k')\hat{a}_s  dk'  \nonumber \\
 &+& \int J_{gf}(x, k'') \hat{\psi}_s(k'')\hat{a}^{\dag}_s  dk'' 
\end{eqnarray} 

with

\begin{align*}
J_f(x,k',k'')& = \frac{i\hbar}{2m}\Bigl( \nabla F^{*}(x,k')F(x,k'') - F^{*}(x,k') \nabla F(x,k'') \Bigr) \\
J_g(x)& = \frac{i\hbar}{2m}\Bigl( \nabla G^{*}(x)G(x) - G^{*}(x) \nabla G(x) \Bigr) \\
J_{fg}(x,k')& = \frac{i\hbar}{2m}\Bigl( \nabla F^{*}(x,k')G(x) - F^{*}(x,k') \nabla G(x) \Bigr) \\
J_{gf}(x,k'')& = \frac{i\hbar}{2m}\Bigl( \nabla G^{*}(x)F(x,k') - G^{*}(x) \nabla F(x,k'') \Bigr) 
\end{align*}
and with $F(x, k') = \frac{1}{\sqrt{2 \pi}}\int f(k, k') e^{ikx}dk$.

\begin{figure}
\includegraphics[width=\columnwidth, bb=0 0 600 600]{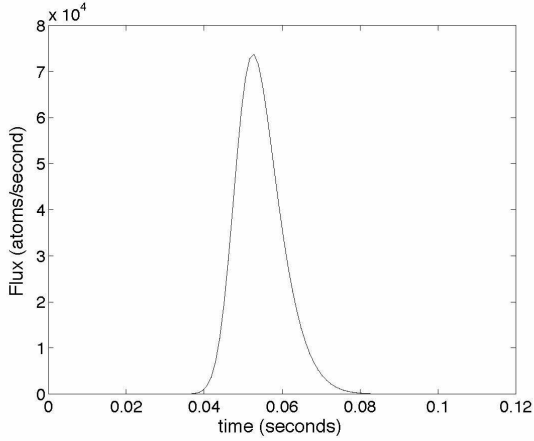}
\caption{\label{fig:flux1} Flux of outcoupled atoms at a point in the atomic beam ($x=1.5$ mm) for $\langle \hat{a}^{\dag}_s \hat{a}_s \rangle = 1000$.}
\end{figure}

Using our initial state $|\psi\rangle = |light\rangle\otimes\{|0\rangle\}_k$, the expectation value of the flux operator becomes:
\begin{equation}
\langle \hat{J}(x)\rangle = J_{g}(x)\langle \hat{a}^{\dag}_s \hat{a}_s \rangle
\end{equation}
This shows that the mean atom flux in the output pulse depends only on the details of the coupling process, and not on the statistics of the outcoupling field.  Figure \ref{fig:flux1} shows the flux of outcoupled atoms for $\langle \hat{a}^{\dag}_s \hat{a}_s \rangle = 1000$. 

To investigate how the quantum statistics are transferred to the atomic beam, we look at the variance in the flux. 

\begin{eqnarray}
V(\hat{J}) &=& \langle \hat{J}^2 \rangle - \langle \hat{J} \rangle^2 \\
&=& J_g^2\langle \hat{a}^{\dag}_s \hat{a}_s \hat{a}^{\dag}_s \hat{a}_s \rangle - J_g^2\langle \hat{a}^{\dag}_s \hat{a}_s\rangle^2 \nonumber \\
 &+& \int\int J_{gf}(x,k')J_{fg}(x,k'') \langle \hat{a}^{\dag}_s \hat{\psi}_s(k') \hat{a}_s \hat{\psi}^{\dag}_s(k'') \rangle dk'dk'' \nonumber \\
 &=& J_g^2 V(\hat{a}^{\dag}_s \hat{a}_s) + \langle \hat{a}^{\dag}_s \hat{a}_s \rangle \int J_{gf}(x,k')J_{fg}(x,k') dk'
\end{eqnarray}
The variance in the flux has two terms, one proportional to the variance in the photon number, and the other proportional to the photon number itself.  For a Fock state photonic input the first of those terms is zero, and the variance is proportional to the function $\int J_{gf}(x,k)J_{fg}(x,k) dk$.  This can be contrasted to the case where the optical field is in a coherent state, and the total variance in the flux is simply proportional to the function $J_g^2 + \int J_{gf}(x,k)J_{fg}(x,k) dk$.  A reasonable measure of the transfer of the quantum state of the zero-dimensional photon field to the larger space of the output pulse is therefore the function
\begin{eqnarray}
v(\hat{J}) &=& \frac{\int J_{gf}(x,k')J_{fg}(x,k') dk'}{J_g^2 + \int J_{gf}(x,k)J_{fg}(x,k) dk},
\end{eqnarray}
which shows the minimum possible variance in the output flux normalised to the flux variance produced by output with a coherent optical state.   

\begin{figure}
\includegraphics[width=\columnwidth, bb=0 0 600 600]{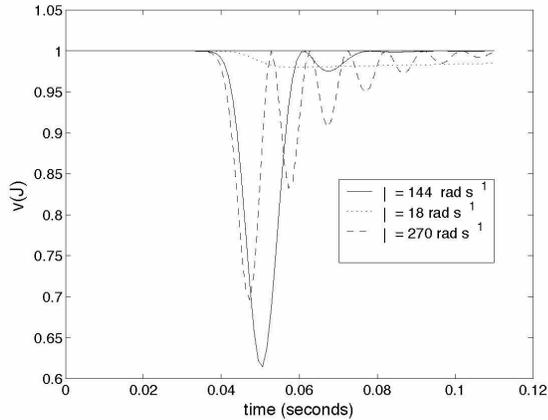}
\caption{\label{fig:fluxfigy} $v(\hat{J})$ at a point in the path of the atomic beam ($x=1.5$ mm) for $\Omega = 18$ rad s$^{-1}$ (dotted line), $\Omega = 144$ rad s$^{-1}$ (solid line), and $\Omega = 270$ rad s$^{-1}$ (dashed line).   Coupling weakly produces a long pulse, but the variance in the flux is almost unaffected by the statistics of the optical state.  Coupling too strongly causes significant back coupling from the output field to the photonic state.}
\end{figure}

\begin{figure}
\includegraphics[width=\columnwidth, bb=0 0 600 600]{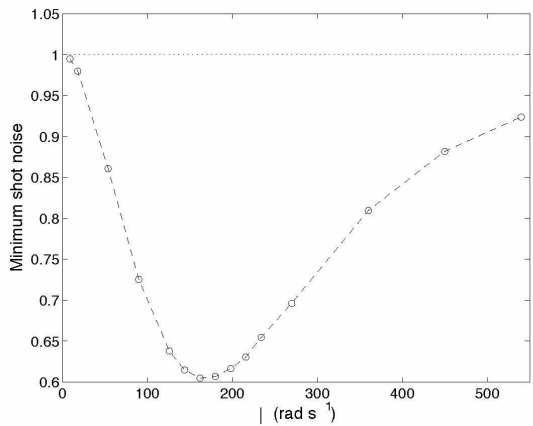}
\caption{\label{fig:maxsqueezy} Minimum value of $v(\hat{J})$ versus $\Omega$ . The shot noise is below the vacuum noise for all values of $\Omega$ when using a Fock state to outcouple. }
\end{figure}

Figure (\ref{fig:fluxfigy}) shows $v(\hat{J})$ for different values of the coupling constant $\Omega$.  Even in our simplified model where we have assumed a single mode for the optical beam and the condensate, the outcoupled atoms still display complicated spatio-temporal dynamics.  Weak outcoupling gives a steady flux, but very little suppression of the shot noise because the timing of the output of each atom becomes uncertain, making the number statistics uncertain in the transient period. When the outcoupling rate is increased, a significant amount of flux squeezing is displayed in a localised pulse. Further increase of the outcoupling rate shows more complicated dynamics, as some of the outcoupled atoms are coupled back into the condensate.  This causes the atoms to come out in a series of pulses, with less flux squeezing than for optimal outcoupling.  An interesting sidenote is that the flux variance produced by the coherent optical state (the denominator of $v(\hat{J})$) is simply proportional to the flux itself, with the same proportionality constant for all times, and all values of $\Omega$. 
Although the variance in the number of the pulse is quite insensitive to the strength of the coupling between the trapped and untrapped fields, this is not true for the variance in the flux.  Figure (\ref{fig:maxsqueezy}) shows the maximum suppression of shot noise in $v(\hat{J})$ for different values of $\Omega$.  We can estimate the outcoupling that will produce the minimum $v(\hat{J})$ by finding the maximum Rabi frequency that will not cause significant back-coupling to the condensate.  Equating the quarter-period of a Rabi oscillation $T_{Rabi}/4 = \pi/(2\Omega)$, with the time taken for the kicked atoms to leave the coupling region $T_{leave} = \sqrt{\frac{8\hbar}{m\omega_{t}}}m/(\hbar|{\bf k}_{23} -{\bf k}_{13}|)$ where $\sqrt{\frac{8\hbar}{m\omega_{t}}}$ is the spatial width of the condensate. From this we can estimate that optimum outcoupling will occur when $\Omega \approx \frac{\pi \hbar|{\bf k}_{23} -{\bf k}_{13}|}{4 m \sqrt{\frac{2\hbar}{m\omega_{t}}}} \approx 250$ rad s$^{-1}$ for the parameters used in this paper.  This agrees well with the calculated minimum shown in figure (\ref{fig:maxsqueezy}).

We have shown that the quantum statistics of an optical mode can be transferred to an atom laser beam to produce a pulse of atoms with better defined number, or to partially suppress the fluctuations in the flux. Furthermore, we have shown that the quantum statistics of the optical mode can be transferred independent of the initial quantum state of the optical mode, which suggests that two-mode optical squeezing could be used to generate spatially separated entangled atomic beams. In the next section we investigate the possibility of using twin optical beams produced from a non-degenerate optical parametric oscillator to generate two entangled atomic beams propagating in different directions.

\section{EPR beams}
\begin{figure}
\includegraphics[width=\columnwidth, bb=0 0 600 600]{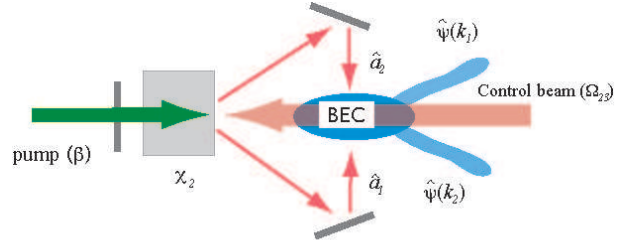}
\caption{\label{fig:eprscheme} Twin atom laser beams produced by outcoupling with two-mode squeezed light. An OPO is driven by a classical, non-depletable driving field ($\beta$). The $\chi^{(2)}$ process produces two optical modes $\hat{a}_1$ and $\hat{a}_2$, which are used to outcouple the atom laser beams. }
\end{figure}

Continuous wave generation of correlated atom beams requires a more complicated scheme.  We consider a probe field created by a nondegenerate OPO producing twin optical beams (figure (\ref{fig:eprscheme})).  These modes have the same wavelength, but travel in different directions and hence they have different momenta. The OPO is driven by a classical, non-depletable driving field. This will produce twin atom laser beams with different momenta. This differs from the previous case in that it allows continuous outcoupling of the atoms, rather than just a pulse. The Hamiltonian for the system is now
\begin{eqnarray}
 \hat{\mathcal H} &=& \hat{\mathcal H}_{atom} \\  \nonumber
 &+& \hbar(\omega-\Delta_1)\hat{a}^{\dag}_1\hat{a}_1 
  + \hbar(\omega-\Delta_1)\hat{a}^{\dag}_2\hat{a}_2 \\ \nonumber
 &+& \hbar\chi(\beta\hat{a}^{\dag}_1\hat{a}^{\dag}_{2} e^{-i\omega_{p}t} +  \beta^{*}\hat{a}_1\hat{a}_{2} e^{i\omega_{p}t}) \\ \nonumber
 &+& \hbar g_{13} \int \hat{\psi}_1(k)\hat{\psi}_3^{\dag}(k+k_1)\hat{a_1} + \hat{\psi}^{\dag}_1(k)\hat{\psi}_3(k+k_1)\hat{a}^{\dag}_1 dk \\ \nonumber
  &+& \hbar g_{13} \int \hat{\psi}_1(k)\hat{\psi}_3^{\dag}(k+k_2)\hat{a_2} + \hat{\psi}^{\dag}_1(k)\hat{\psi}_3(k+k_2)\hat{a}^{\dag}_2 dk \\ \nonumber
&+& \hbar \int \Omega_{23} \hat{\psi}^{\dag}_2(k)\hat{\psi}_3(k+k_0)e^{i(\omega-\Delta_2)t} \\ \nonumber
&+&  \Omega^{*}_{23} \hat{\psi}_2(k)\hat{\psi}^{\dag}_3(k+k_0)e^{-i(\omega-\Delta_2)t} dk \\ \nonumber
 \end{eqnarray}
where $\hat{a}_1$ and $\hat{a}_2$ represent the annihilation operators for the twin probe fields produced from the down conversion process, both assumed to affect the $|1\rangle \rightarrow |3\rangle$ transition, $\beta$ is the complex amplitude of the pump field, $\omega_p$ is the frequency of the pump,  and $\chi$ is the nonlinear coefficient of the down conversion medium. $\hbar k_1$ and $\hbar k_2$ are the magnitudes of the momentum kicks due to absorption from photons in $\hat{a}_1$ and $\hat{a}_2$ respectively. We have assumed that the photons are resonant in an optical resonator with 100\% reflective mirrors. This assumption is valid as the dominant form of loss out of the cavity will be due to atomic absorption.  For computational convenience in our one-dimensional model, we have chosen $\mathbf{k_1}-\mathbf{k_0} = -(\mathbf{k_2}-\mathbf{k_0})$, ie. the resultant momentum kicks that the atoms obtain after being outcoupled are of equal magnitude and opposite direction.  By adiabatically eliminating the excited state, and assuming the condensate is a large coherent state as before, we obtain the following equations of motion for the outcoupled atoms and the probe fields:
\begin{eqnarray}  \label{OPOEOM}
i\dot{\hat{\psi}}(k) &=& \omega_0(k)\hat{\psi}(k) - \Omega_1(k)\tilde{a}_1  + \Omega_2(k)\tilde{a}_2 \\
i\dot{\tilde{a}}_1 &=& \omega_a \tilde{a}_1  - \int \Omega_1^{*}(k)\hat{\psi}(k)dk  \\ \nonumber
&+& \chi\beta\tilde{a}^{\dag}_2 e^{i(2(\omega -\Delta_2)-\omega_{p})t} -\Omega_C \tilde{a}_2 \\
i\dot{\tilde{a}}_2 &=& \omega_a \tilde{a}_2  - \int \Omega_2^{*}(k)\hat{\psi}(k)dk  \\ \nonumber
 &+& \chi\beta\tilde{a}^{\dag}_1 e^{i(2(\omega -\Delta_2)-\omega_{p})t} -\Omega_C^{*} \hat{a}_1
\end{eqnarray}
 with $\hat{\psi}(k) = \hat{\psi}_2(k)e^{i\omega_{t}t}$, $\tilde{a}_j = \hat{a}_j e^{i(\omega-\Delta_2)t}$, $\Omega_j(k) =  g\sqrt{N} \frac{\Omega}{\Delta_2} \phi_0(k+k_0-k_j)$ for $j=1, 2$, and $\Omega_C = \frac{g^2N}{\Delta_2}\int\phi_0^{*}(k-k_1)\phi_0(k-k_2) dk$. The $\Omega_C$ cross coupling term between the two optical modes is due to atoms absorbing a photon from one beam and emitting it into the other beam. This term will be small due to the large momentum difference between the two modes.  However, the functional form of $\Omega_C$ is due to our assumption that the condensate remains single mode. Cross coupling between the two optical modes will cause momentum `side bands' on the condensate mode \cite{Moore}, but the effect of this cross coupling will be small if the number of photons in the probe beam is small compared to the number of atoms in the condensate.  As the results in this section are calculated in a parameter regime where the chance of an outcoupled atom coupling back into the condensate is low, it is valid to neglect this term in our calculations. The general solution to equations (\ref{OPOEOM}) is
 \begin{eqnarray} \label{oposol}
 \hat{\psi}(k,t) &=& \int f_{+}(k,k',t)\hat{\psi}_s(k)dk' \\ \nonumber
  &+& \int f_{-}(k,k',t)\hat{\psi}^{\dag}_s(k)dk'+ g_{1+}(k,t)\hat{a}_{1s} \\ \nonumber &+& g_{1-}(k,t)\hat{a}^{\dag}_{1s} 
  + g_{2+}(k,t)\hat{a}_{2s} + g_{2-}(k,t)\hat{a}^{\dag}_{2s} \\
\hat{a}_1(t) &=& p_{1+}(t)\hat{a}_{1s} + p_{1-}(t)\hat{a}^{\dag}_{1s} \\ \nonumber
 &+& p_{2+}(t)\hat{a}_{2s} + p_{2-}(t)\hat{a}^{\dag}_{2s} \\ \nonumber
 &+&  \int p_{3+}(k',t)\hat{\psi}_s(k')dk' +  \int p_{3-}(k',t)\hat{\psi}^{\dag}_s(k')dk' \\ 
 \hat{a}_2(t) &=& q_{1+}(t)\hat{a}_{1s} + q_{1-}(t)\hat{a}^{\dag}_{1s} \\ \nonumber
  &+& q_{2+}(t)\hat{a}_{2s} + q_{2-}(t)\hat{a}^{\dag}_{2s} \\ \nonumber
 &+&  \int q_{3+}(k',t)\hat{\psi}_s(k')dk' +  \int q_{3-}(k',t)\hat{\psi}^{\dag}_s(k')dk'
\end{eqnarray}
where $f_{\pm}(k,k',t)$, $g_{1,2 \pm}(k,t)$, $p_{1,2\pm}(t)$, $p_{3\pm}(k',t)$, $q_{1,2 \pm}(t)$, $q_{3\pm}(k',t)$ are complex functions satisfying differential equations obtained by substituting the solutions (\ref{oposol}) into (\ref{OPOEOM}) (see appendix). From the solution of these equations, we can calculate any observable of the system.

\begin{figure}
\includegraphics[width=\columnwidth, bb=0 0 600 600]{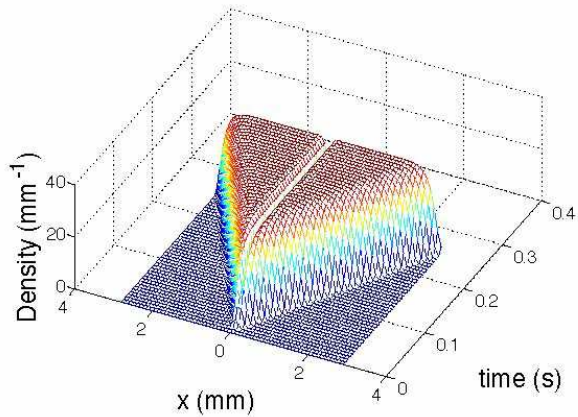}
\caption{\label{fig:opodensity} Density of outcoupled atoms versus time. Outcoupling using light from a non-degenerate OPO produces two atomic beams in opposite directions, with steady flux. }
\end{figure}

\begin{figure}
\includegraphics[width=\columnwidth, bb=0 0 600 600]{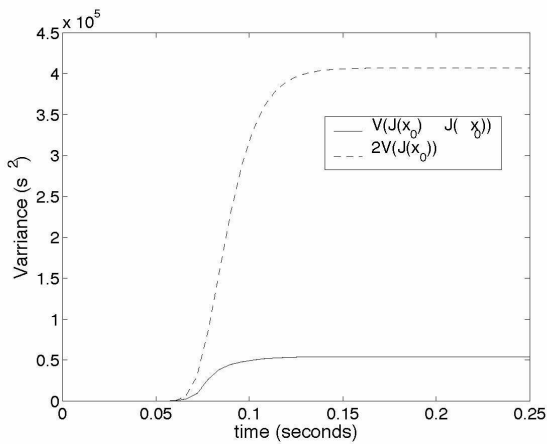}
\caption{\label{fig:opovariance} Variance in the flux difference. The fluctuations are eight times smaller than for uncorrelated atoms. }
\end{figure}

We solved the equations (\ref{opoprop}) numerically for $\chi\beta = 80$s$^{-1}$, $\Omega_j(k) = \Omega\phi_0(k-k_0-k_j)$ with $\Omega = 108$ rad s$^{-1}$ for $j=1, 2$. We set $k_1-k_0 = -(k_2 -k_0) = 1.6\times10^7$ m$^{-1}$, and $\omega_p- 2(\omega-\Delta_2) = 2\omega_a$, with all other parameters as before. If we assume that the two optical modes and the untrapped atomic field are initially in the vacuum state, using equation [\ref{oposol}] the expectation value of the atomic density $\rho(x) = \langle \hat{\Psi}^{\dag}(x)\hat{\Psi}(x) \rangle$ is
\begin{eqnarray}
 \rho(x) &=& \langle \hat{\Psi}^{\dag}(x)\hat{\Psi}(x) \rangle \\ \nonumber
             &=& \int |F_{-}(x, k')|^2 dk' + |G_{1-}(x)|^2 + |G_{2-}(x)|^2 \nonumber
\end{eqnarray}
Where $F_{-}(x, k')=\frac{1}{\sqrt{2 \pi}}\int f_{-}(k, k') e^{ikx}dk$, $G_{1-}(x)=\frac{1}{\sqrt{2 \pi}}\int g_{1-}(k) e^{ikx}dk $ and $G_{2-}(x)=\frac{1}{\sqrt{2 \pi}}\int g_{2-}(k) e^{ikx}dk $. Figure(\ref{fig:opodensity}) shows the atomic density versus time. Two atomic beams in opposite directions are produced, with steady flux. 
 To check whether there are correlations present in the two atomic beams, the relevant observable is the difference in the flux of the two beams. If the two beams are completely uncorrelated, then we would expect
\begin{equation}
V(\hat{J}(x_0) - \hat{J}(-x_0)) \geq 2V(\hat{J}(x_0)) \label{fluxsqueezecondition}
\end{equation} 
 Where $x_0$ and $-x_0$ are points that lie in the rightward and leftward propagating beams respectively. Figure (\ref{fig:opovariance}) shows the variance of the flux difference $V(\hat{J}(x_0) - \hat{J}(-x_0))$ versus time, and shows that the fluctuations in the difference are approximately eight times less than for uncorrelated atoms. 

\begin{figure}
\includegraphics[width=\columnwidth, bb=0 0 600 600]{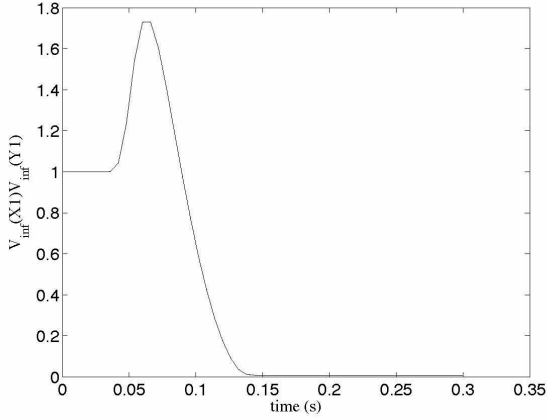}
\caption{\label{fig:epr1} Product of the inferred variances $V^{inf}(\hat{X}_{-})V^{inf}(\hat{Y}_{-})$ versus time. As the system goes to steady state, the requirement for entanglement is satisfied.}
\end{figure}

We now investigate whether we can use this process to generate entanglement between the two atomic beams. We quantify our entanglement using the EPR criterion of Reid and Drummond \cite{Reid}, the requirement being that two conjugate variables on one of the beams can be inferred from measurements on the other beam to below the quantum limit. We define four quadratures:
\begin{eqnarray}
\hat{X}_{\pm} &=& \int(L_{\pm}^{*}(x)\hat{\Psi}(x) + L_{\pm}(x)\hat{\Psi}^{\dag}(x))dx \\ 
\hat{Y}_{\pm} &=& i\int(L_{\pm}^{*}(x)\hat{\Psi}(x) - L_{\pm}(x)\hat{\Psi}^{\dag}(x))dx \\ 
\end{eqnarray}
with
\begin{eqnarray}
 L_{\pm}(x,t) &=& \frac{e^{i(k_0x-\omega_a t)}}{\sqrt{|x_1-x_2|}},\phantom{aa}  \mbox{if} \pm x_1 > x > \pm x_2 \nonumber \\
   &=& 0 \quad \mbox{otherwise}
\end{eqnarray}
The commutator of the conjugate quadratures give us the uncertainty relation $V(\hat{X}_{\pm})V(\hat{Y}_{\pm}) \geq 1$ since $\int |L_{\pm}(x)|^2dx =1$. The beams are entangled under the EPR criterion if by making measurements of quadratures of one beam (eg. $\hat{X}_{+}$ and $\hat{Y}_{+}$), then quadratures of the other beam ($\hat{X}_{-}$ and $\hat{Y}_{-}$) can be inferred to better than this quantum limit. Quantitatively: $V^{inf}(\hat{X}_{-})V^{inf}(\hat{Y}_{-}) < 1$  is the requirement for entanglement, where $V^{inf}(\hat{X}_{\pm}) = V(\hat{X}_{\pm}) - \frac{(V(\hat{X}_{\pm}, \hat{Y}_{\mp}))^2}{V(\hat{Y}_{\mp})}$,  $V^{inf}(\hat{Y}_{\pm}) = V(\hat{Y}_{\pm}) - \frac{(V(\hat{Y}_{\pm}, \hat{X}_{\mp}))^2}{V(\hat{X}_{\mp})}$ and $V(a,b) = \langle ab\rangle - \langle a\rangle\langle b\rangle$. We note here that the correlations present are between conjugate quadratures of each beam. Figure (\ref{fig:epr1}) shows the product of the inferred variances $V^{inf}(\hat{X}_{-})V^{inf}(\hat{Y}_{-})$ plotted against time. As the intensity of the beams increase and become more monochromatic, the product of the inferred variances dip well below the  requirement for entanglement. The initial increase is due to the beams initially not approximating plane waves. This could be fixed by appropriate choice of $L_{\pm}(x)$, to better match the mode shape of the output atom laser beams. 
In the long time limit, the product of the inferred variances is on the order of three orders of magnitude below the classical limit, demonstrating that this system produces an almost pure EPR correlated state.  Our model uses an ideal OPO, so the squeezing in the optical modes would grow without bound in the absence of the damping due to the atoms.  In practice, the entanglement on the atomic beams will not exceed the optical entanglement that can be obtained from a real OPO.  The limit to the entanglement between the atomic beams in this model is given by the finite momentum width of the condensate.  As the EPR state is expected to be very pure, the dominant noise in an experiment may actually be due to some of the effects we have ignored in our model due to their small effect on the dynamics.  In particular, there may be a small reduction in fidelity due to effects of the back action of the outcoupling on the condensate wavefunction, which we have ignored in this calculation.

\section{Conclusion}
We have modelled the dynamics of an atom laser produced by outcoupling from a Bose-Einstein condensate with squeezed light. We modelled the multimode dynamics of the output field and showed that a significant amount of squeezing can be transferred from an optical mode to a propagating atom laser beam.  We also demonstrated that two-mode squeezing can be used to produce twin atom laser beams with continuous variable entanglement in amplitude and phase.

This research was supported by the Australian Research Council Centre of Excellence for Quantum Atom Optics. We would like to acknowledge useful discussions with M. K. Olsen, A. M. Lance and H. A. Bachor. 

\section{Appendix}
$f_{+}(k,k',t)$, $f_{-}(k,k',t)$, $g_{1+}(k,t)$, $g_{1-}(k,t)$, $g_{2+}(k,t)$, $g_{2-}(k,t)$, $p_{1+}(t)$, $p_{1-}(t)$, $p_{2+}(t)$, $p_{2-}(t)$, $p_{3+}(k', t)$, $p_{3-}(k', t)$, $q_{1+}(t)$, $q_{1-}(t)$, $q_{2+}(t)$, $q_{2-}(t)$, $q_{3+}(k', t)$, $q_{3-}(k', t)$, must satisfy:
\begin{eqnarray} \label{opoprop}
i \dot{f}_{+}(k, k') &=& \omega_0(k)f_{+}(k, k') - \Omega_1(k)p_{3+}(k') \\ \nonumber
  &-& \Omega_2(k)q_{3+}(k') \\  \nonumber
i \dot{f}_{-}(k, k') &=& \omega_0(k)f_{-}(k, k') - \Omega_1(k)p_{3-}(k') - \Omega_2(k)q_{3-}(k') \\ \nonumber
i \dot{g}_{1+}(k) &=& \omega_0(k)g_{1+}(k) - \Omega_1(k)p_{1+} - \Omega_2(k)q_{1+} \\ \nonumber
i \dot{g}_{1-}(k) &=& \omega_0(k)g_{1-}(k) - \Omega_1(k)p_{1-} - \Omega_2(k)q_{1-} \\ \nonumber
i \dot{g}_{2+}(k) &=& \omega_0(k)g_{2+}(k) - \Omega_1(k)p_{2+} - \Omega_2(k)q_{2+} \\ \nonumber
i \dot{g}_{2-}(k) &=& \omega_0(k)g_{2-}(k) - \Omega_1(k)p_{2-} - \Omega_2(k)q_{2-} \\ \nonumber
i\dot{p}_{1+} &=& \omega_a p_{1+} - \int\Omega_1^{*}(k)g_{1+}(k)dk +\chi_{p}q_{1-}^{*} \\ \nonumber
i\dot{p}_{1-} &=& \omega_a p_{1-} - \int\Omega_1^{*}(k)g_{1-}(k)dk +\chi_{p}q_{1+}^{*} \\ \nonumber
i\dot{p}_{2+} &=& \omega_a p_{2+} - \int\Omega_1^{*}(k)g_{2+}(k)dk +\chi_{p}q_{2-}^{*} \\ \nonumber
i\dot{p}_{2-} &=& \omega_a p_{2-} - \int\Omega_1^{*}(k)g_{2-}(k)dk +\chi_{p}q_{2+}^{*} \\ \nonumber
i\dot{p}_{3+}(k') &=& \omega_a p_{3+}(k') - \int\Omega_1^{*}(k)f_{+}(k, k')dk +\chi_{p}q_{3-}^{*}(k') \\ \nonumber
i\dot{p}_{3-}(k') &=& \omega_a p_{3-}(k') - \int\Omega_1^{*}(k)f_{-}(k, k')dk +\chi_{p}q_{3+}^{*}(k') \\ \nonumber
i\dot{q}_{1+} &=& \omega_a q_{1+} - \int\Omega_1^{*}(k)g_{1+}(k)dk +\chi_{p}p_{1-}^{*} \\ \nonumber
i\dot{q}_{1-} &=& \omega_a q_{1-} - \int\Omega_1^{*}(k)g_{1-}(k)dk +\chi_{p}p_{1+}^{*} \\ \nonumber
i\dot{q}_{2+} &=& \omega_a q_{2+} - \int\Omega_1^{*}(k)g_{2+}(k)dk +\chi_{p}p_{2-}^{*} \\ \nonumber
i\dot{q}_{2-} &=& \omega_a q_{2-} - \int\Omega_1^{*}(k)g_{2-}(k)dk +\chi_{p}p_{2+}^{*} \\ \nonumber
i\dot{q}_{3+}(k') &=& \omega_a q_{3+}(k') - \int\Omega_1^{*}(k)f_{+}(k, k')dk +\chi_{p}p_{3-}^{*}(k') \\ \nonumber
i\dot{q}_{3-}(k') &=& \omega_a q_{3-}(k') - \int\Omega_1^{*}(k)f_{-}(k, k')dk +\chi_{p}p_{3+}^{*}(k') \nonumber 
\end{eqnarray}
with $\chi_{p} = \chi\beta e^{i(2(\omega -\Delta_2)-\omega_{p})t}$, and initial conditions $f_{+}(k, k', t=0) = \delta(k-k')$, $p_{1+}(t=0) = 1$, $q_{2+}(t=0) = 1$ with all other fields zero. From the solution of these equations, we can calculate any observable of the system.

\end{document}